\documentclass[prd,twocolumn,showpacs,floatfix,amsmath,amssymb,floatfix,nofootinbib]{revtex4}
\usepackage{graphicx,color,dcolumn,booktabs,bm}
\usepackage{amsmath,amssymb,amsfonts,slashed}
\usepackage{longtable,lscape,comment,braket}
\usepackage{diagbox}
\usepackage{txfonts}
\usepackage[mathscr]{euscript}
\usepackage{overpic}
\usepackage{indentfirst}
\usepackage{feynmf}   
\usepackage{slashed}  
\usepackage{cases}
\usepackage{epstopdf}
\usepackage{psfrag}
\usepackage{subfigure}
\usepackage{threeparttable}
\pdfoutput=1
\usepackage{color}
\usepackage{multirow}

\graphicspath{{Figures/}}
\usepackage[colorlinks, citecolor=blue,anchorcolor=red,menucolor=red,
linkcolor=red,filecolor=red,runcolor=red,urlcolor=blue,frenchlinks=red]{hyperref}
\usepackage{ulem}

\def\bal#1\eal{\begin{align}#1{}\end{align}}
\def\cce/{coupled-channel effects}
\newcommand{\vect}[1]{\boldsymbol{\mathbf{#1}}}
\newcommand \inner [2] {\langle {#1}\vert {#2}\rangle}

\def\be{\begin{equation}}
\def\ee{\end{equation}}
\def\ba{\begin{eqnarray}}
\def\ea{\end{eqnarray}}
\def\3P0{{}^3P_0}

\begin{document}

\title{$\chi_{b}(3P)$ Multiplet Revisited: Ultrafine Mass Splitting and Radiative Transitions}

\author{Muhammad Naeem Anwar$^{1,2,3}$}\email[]{m.anwar@fz-juelich.de}
\author{Yu Lu$^{4}$}\email[]{luyu@hiskp.uni-bonn.de}
\author{Bing-Song Zou$^{1,2}$}\email[]{zoubs@itp.ac.cn}
\affiliation{$^1$CAS Key Laboratory of Theoretical Physics, Institute of Theoretical Physics, Chinese Academy of Sciences, Beijing 100190, China\\
$^2$University of Chinese Academy of Sciences, Beijing 100049, China\\
$^3$Institute for Advanced Simulation, Institut f\"ur Kernphysik and J\"ulich Center for Hadron Physics, Forschungszentrum J\"ulich, D-52425 J\"ulich, Germany\\
$^4$Helmholtz-Institut f\"ur Strahlen- und Kernphysik and Bethe Center for Theoretical Physics, Universit\"at Bonn,  D-53115 Bonn, Germany}

\date{\today}

\begin{abstract}

Invoked by the recent CMS observation regarding candidates of the $\chi_b(3P)$ multiplet, we analyze the ultrafine and mass splittings among $3P$ multiplet in our unquenched quark model (UQM) studies. The mass difference of $\chi_{b2}$ and $\chi_{b1}$ in $3P$ multiplet measured by CMS collaboration ($10.6 \pm 0.64 \pm 0.17$ MeV) is very close to our theoretical prediction ($12$ MeV). Our corresponding mass splitting of $\chi_{b1}$ and $\chi_{b0}$ enables us to predict more precisely the mass of $\chi_{b0}(3P)$ to be ($10490\pm 3$) MeV. Moreover, we predict ratios of the radiative decays of $\chi_{bJ}(nP)$ candidates, both in UQM and quark potential model. Our predicted relative branching fraction of $\chi_{b0}(3P)\to\Upsilon(3S)\gamma$ is one order of magnitude smaller than $\chi_{b2}(3P)$, this naturally explains the non-observation of $\chi_{b0}(3P)$ in recent CMS search. We hope these results might provide useful references for forthcoming experimental searches.

\end{abstract}

\maketitle

\section{Introduction}

The excited $P$-wave bottomonia, $\chi_{bJ}(3P)$, are of special interest, since they provide a laboratory to test (and model) the non-perturbative spin-spin interactions of heavy quarks. Very recently, the CMS collaboration observed two candidates of the bottomonium $3P$ multiplet, $\chi_{b1}(3P)$ and $\chi_{b2}(3P)$, through their decays into $\Upsilon(3S) \gamma$~\cite{Sirunyan:2018dff}. Their measured masses and mass splitting are
\be
\begin{split}
M[\chi_{b1}(3P)]=(10513.42 \pm 0.41 \pm 0.18)~\mathrm{MeV}~,\\
M[\chi_{b2}(3P)]=(10524.02 \pm 0.57 \pm 0.18)~\mathrm{MeV}~,\\
\Delta m_{21} \equiv m(\chi_{b2})-m(\chi_{b1})= (10.6 \pm 0.64 \pm 0.17)~\mathrm{MeV}~.
\end{split}
\ee

There are some earlier measurements related to $\chi_{bJ}(3P)$ mass by ATLAS~\cite{Aad:2011ih}, LHCb~\cite{Aaij:2014hla,Aaij:2014caa}, and D0 Collaborations~\cite{Abazov:2012gh}. However, these measurements can not distinguish between the candidates of $\chi_{bJ}(3P)$ multiplet. The recent CMS analysis~\cite{Sirunyan:2018dff} is higher resolution search, and hence, is able to distinguish between $\chi_{b1}(3P)$ and $\chi_{b2}(3P)$ for the first time.

In this paper we intend to compare our unquenched quark model studies with this recent measurement, and make more precise prediction for the mass of the other $3P$ bottomonium ($\chi_{b0}$) by incorporating the measured mass splitting. We also make an analysis of the ultrafine splitting of $P$-wave bottomonia, which enlighten the internal quark structure of the considered bottomonium. In addition, we predict model-independent ratios of radiative decays of $\chi_{bJ}(nP)$ candidates.

Heavy quarkonium states can couple to intermediate heavy mesons through
the creation of light quark-antiquark pair which enlarge the Fock space
of the initial state, i.e. the initial state contains multiquark components.
These multiquark components will change the Hamiltonian of the potential model,
causing the mass shift and mixing between states with the same quantum numbers
or directly contributing to open channel strong decay if the initial state is above threshold.
These can be summarized as coupled-channel effects (CCE). When CCE are combined with
the naive quark potential model, one gets the unquenched quark model (UQM).
UQM has been considered at least 35 years ago by
T\"{o}rnqvist \textit{et al.}~\cite{Heikkila:1983wd,Ono:1983rd,Ono:1985jt,Ono:1985eu}.

The physical or experimentally observed bottomonium state $\ket{A} $ is expressed in UQM as
\be
\ket {A}=c_0 \ket{\psi_0} +\sum_{BC} \int d^3p\, c_{BC}(p) \ket{BC;p},
\ee
where $c_0$ and $c_{BC}$ stand for the normalization constants of the bare state and the $BC$ components, respectively.
In this work, $B$ and $C$ refer to bottom and anti-bottom mesons,
and the summation over $BC$ is carried out including all possible pairs of ground-state bottom mesons.
The $\ket {\psi_0}$ is normalized to 1 and $\ket {A}$ is also normalized to 1 if it lies below $B\bar{B}$ threshold,
and $\ket{BC;p}$ is normalized as $\inner{BC;p_1}{B'C';p_2}=\delta^3(p_1-p_2)\delta_{BB'}\delta_{CC'}$,
where $p$ is the momentum of $B$ meson in $\ket{A}$'s rest frame. The full Hamiltonian of the
physical state then reads as
\be
  H=H_0+H_{BC}+H_I,
\ee
where $H_0$ is the Hamiltonian of the bare state (see Appendix~\ref{bare} for details), $H_{BC} \ket{BC;p} = E_{BC}\ket{BC;p}$ with
$E_{BC}=\sqrt{m_B^2+p^2}+\sqrt{m_C^2+p^2}$ is the energy of the continuum state
(interaction between $B$ and $C$ is neglected and the transition between one continuum to another is restricted), and $H_I$
is the interaction Hamiltonian which mix the bare state with the continuum.
Since each quark pair creation model generates its own vertex functions that in turn lead
to specific real parts of hadronic loops, see Ref.~\cite{Hammer:2016prh} for related remarks.

\begin{figure}[h]
  \centering
  \includegraphics[width=0.5\textwidth]{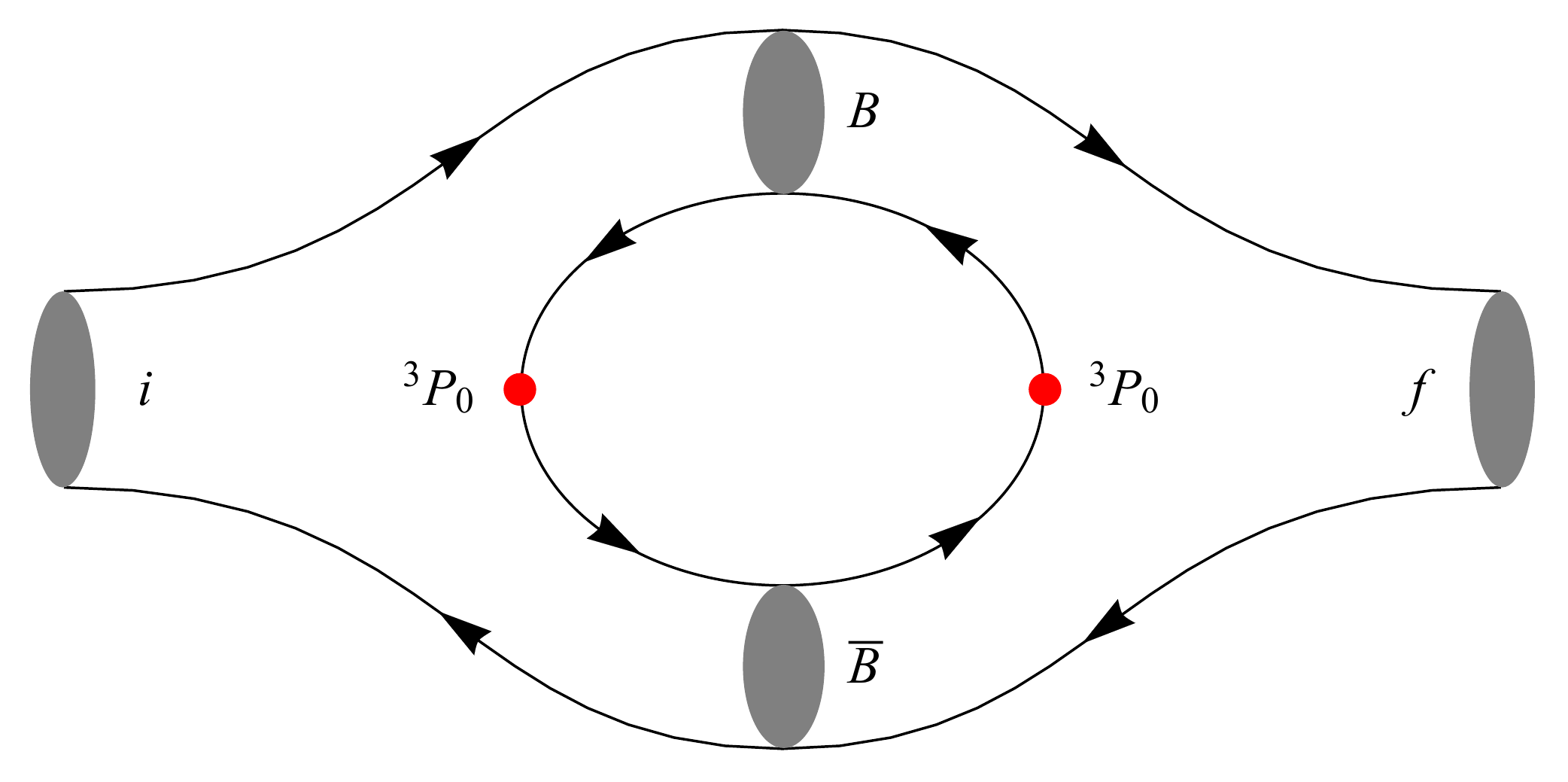}\\
  \caption{Sketch of coupled-channel effects in $\3P0$ model.
   $i$ and $f$ respectively denote the initial and final states with same $J^{PC}$ and
   $B\bar{B}$ stands for all possible $B$ meson pairs.
  }
 \label{ccDiagram}
\end{figure}

Here, for the bare-continuum mixing, we adopt the widely used ${}^3P_0$ model~\cite{Micu:1968mk}. In this model, the generated
quark pairs have vacuum quantum numbers $J^{PC}=0^{++}$ which in spectroscopical notation
${}^{2S+1}L_J$ equals to $\3P0$. A sketch of ${}^3P_0$ model induced mixing is shown in Fig.~\ref{ccDiagram}.
The interaction Hamiltonian can be expressed as
\be
H_I=2 m_q \gamma \int d^3x \bar{\psi}_q \psi_q,
\ee
where $m_q$ is the produced quark mass,
and $\gamma$ is the dimensionless coupling constant.
The $\psi_q$ ($\bar{\psi}_q$) is the spinor field to generate anti-quark (quark).
Since the probability to generate heavier quarks is suppressed,
we use the effective strength $\gamma_s=\frac{m_q}{m_s}\gamma$ in the following calculation,
where $m_q=m_u=m_d$ is the constituent quark mass of up (or down) quark and $m_s$  is strange quark mass.

The mass shift caused by the $BC$ components and the probabilities of the $b\bar{b}$ core are obtained
after solving the Schr\"{o}dinger equation with the full Hamiltonian $H$.
They are expressed as
\bal
\Delta M&:=M-M_0=\sum_{BC}\int d^3p\, \frac{\vert \bra {BC;p} H_I \ket{\psi_0} \vert ^2}{M-E_{BC}-i\epsilon}, \label{massShift}\\
P_{b\bar{b}}&:=\vert c_0 \vert ^2=\left(1+\sum_{BC LS}\int dp \frac{p^2 \bra {BC;p} H_I \ket{\psi_0} \vert ^2}{(M-E_{BC})^2}\right)^{-1}, \label{prob}
\eal
where $M$ and $M_0$ are the eigenvalues of the full ($H$) and quenched/bare Hamiltonian ($H_0$), respectively.
See Appendix~\ref{prob} or Refs.~\cite{Lu:2016mbb,Ferretti:2013vua} for derivation of above relations and UQM calculation details.
Numerical values of $\Delta M$ and $P_{b\bar{b}}$ of every coupled channel for the bottomonia below $B\bar{B}$ threshold
are given in Table~\ref{tab:massProb}, which will be used in the following discussions.

\begin{table*}[!htbp]
  \renewcommand\arraystretch{1.5}
  \setlength{\tabcolsep}{4pt}
    \centering
  \begin{tabular}{c|cc|cc|cc|cc|cc|cc|cc}
    \hline\hline
   Initial States
   &\multicolumn{2}{c|}{$B\bar{B}$}
   &\multicolumn{2}{c|}{$B\bar{B}^*+h.c.$}
   &\multicolumn{2}{c|}{$B^*\bar{B}^*$}
   &\multicolumn{2}{c|}{$B_s\bar{B}_s$}
   &\multicolumn{2}{c|}{$B_s\bar{B}_s^*+h.c.$}
   &\multicolumn{2}{c|}{$B_s^*\bar{B}^*_s$}
   &\multicolumn{2}{c}{Total}
   \\ \cline{2-15}
   &$-\Delta M$&$P_{b\bar{b}}$
   &$-\Delta M$&$P_{b\bar{b}}$
   &$-\Delta M$&$P_{b\bar{b}}$
   &$-\Delta M$&$P_{b\bar{b}}$
   &$-\Delta M$&$P_{b\bar{b}}$
   &$-\Delta M$&$P_{b\bar{b}}$
   &$-\Delta M$&$P_{b\bar{b}}(\%)$
   \\ \hline
   $\eta_b(1S)$ & 0 & 0 & 7.8  & 0.45 & 7.6  & 0.43 & 0 & 0 & 3.3 & 0.17 & 3.3 & 0.16 & 22.0 & 98.79 \\
   $\eta_b(2S)$ & 0 & 0 & 16.5 & 1.81 & 15.7 & 1.62 & 0 & 0 & 5.2 & 0.43 & 5.0 & 0.4  & 42.4 & 95.74 \\
   $\eta_b(3S)$ & 0 & 0 & 24.5 & 5.01 & 22.3 & 3.98 & 0 & 0 & 5.4 & 0.63 & 5.1 & 0.55 & 57.4 & 89.83 \\
   \hline
   $\Upsilon(1S)$ & 1.4 & 0.09 & 5.4  & 0.33 & 9.2  & 0.54 & 0.6 & 0.03 & 2.3 & 0.12 & 3.9 & 0.2  & 22.8 & 98.69 \\
   $\Upsilon(2S)$ & 3.0 & 0.37 & 11.4 & 1.29 & 18.9 & 2.02 & 0.9 & 0.08 & 3.5 & 0.31 & 5.9 & 0.49 & 43.8 & 95.44 \\
   $\Upsilon(3S)$ & 4.8 & 1.25 & 17.2 & 3.71 & 27.1 & 5.07 & 1.0 & 0.13 & 3.7 & 0.45 & 6.1 & 0.67 & 60.0 & 88.71 \\
    \hline
   $h_b(1P)$ & 0 & 0 & 13.5 & 1.22  & 13.0 & 1.12 & 0 & 0 & 4.8 & 0.35 & 4.6 & 0.33 & 35.8 & 96.99 \\
   $h_b(2P)$ & 0 & 0 & 21.9 & 3.51  & 20.3 & 2.96 & 0 & 0 & 5.6 & 0.59 & 5.3 & 0.52 & 53.1 & 92.43 \\
   $h_b(3P)$ & 0 & 0 & 38.0 & 19.75 & 29.5 & 9.04 & 0 & 0 & 5.4 & 0.67 & 5.0 & 0.54 & 77.9 & 70.0  \\
    \hline
   $\chi_{b0}(1P)$ & 4.1  & 0.45  & 0 & 0 & 21.4 & 1.74 & 1.3 & 0.11 & 0 & 0 & 7.8 & 0.52 & 34.6 & 97.18 \\
   $\chi_{b0}(2P)$ & 9.3  & 1.85  & 0 & 0 & 31.1 & 4.13 & 2.1 & 0.26 & 0 & 0 & 8.4 & 0.77 & 50.9 & 92.98 \\
   $\chi_{b0}(3P)$ & 25.5 & 34.08 & 0 & 0 & 40.7 & 8.07 & 2.3 & 0.31 & 0 & 0 & 7.6 & 0.62 & 76.1 & 56.92 \\
    \hline
   $\chi_{b1}(1P)$ & 0 & 0 & 10.8 & 1.03 & 15.5 & 1.27 & 0 & 0 & 3.7 & 0.28 & 5.6 & 0.38 & 35.5 & 97.03 \\
   $\chi_{b1}(2P)$ & 0 & 0 & 19.7 & 3.38 & 22.1 & 3.0  & 0 & 0 & 4.8 & 0.53 & 6.0 & 0.56 & 52.6 & 92.53 \\
   $\chi_{b1}(3P)$ & 0 & 0 & 37.4 & 21.9 & 29.7 & 7.54 & 0 & 0 & 4.8 & 0.64 & 5.4 & 0.54 & 77.4 & 69.38 \\
    \hline
   $\chi_{b2}(1P)$ & 3.4  & 0.31 & 9.8  & 0.85  & 13.6 & 1.24  & 1.2 & 0.09 & 3.5 & 0.25 & 4.7 & 0.35 & 36.4 & 96.91 \\
   $\chi_{b2}(2P)$ & 5.3  & 0.89 & 14.6 & 2.23  & 23.2 & 3.62  & 1.3 & 0.15 & 3.8 & 0.39 & 5.8 & 0.6  & 54.1 & 92.13 \\
   $\chi_{b2}(3P)$ & 12.3 &  --  & 23.3 & 12.50 & 36.2 & 16.34 & 1.3 & 0.23 & 3.6 & 0.53 & 5.6 & 0.82 & 82.2 & 69.57 \\
    \hline \hline
  \end{tabular}
  \caption{The mass shift (in MeV) and probability (in \%) of every coupled channel for the bottomonia below $B\bar{B}$ threshold.
Note that since $h_b(3P)$ has no coupling to $B\bar{B}$,
even though $h_b(3P)$ is above $B\bar{B}$ threshold,
the probability is still well-defined.
However, since $\chi_{b2}(3P)$ couples to $B\bar{B}$ channel and lies above this threshold,
causing difficulty to the renormalization of the wave function.
We make the assumption that the renormalization caused by $B\bar{B}$ channel can be discarded,
see Sec.~\ref{trans} for related discussions. 
}
\label{tab:massProb}
\end{table*}

\section{Mass Splitting and $\chi_{b0}(3P)$}

After the recent CMS observation~\cite{Sirunyan:2018dff} of $\chi_{b1}(3P)$ and $\chi_{b2}(3P)$, $\chi_{b0}(3P)$ is now the only missing candidate in spin-triplet $3P$ bottomonium. With the reference of observed mass splitting of $1P$, $2P$ and $3P$ multiplets, one can predict the mass of $\chi_{b0}(3P)$. It requires a constraint that the mass splittings for $1P$, $2P$ and $3P$ multiplet should be the same~\cite{Dib:2012vw}.

Triggered by the above mentioned experimental search, we analyze our UQM studies regarding the bottomonium spectrum~\cite{Lu:2016mbb,Lu:2017hma}. We notice that the measured mass splitting between $\chi_{b2}(3P)$ and $\chi_{b1}(3P)$ is $(10.6 \pm 0.64 \pm 0.17)$ MeV which differs only by $1$ MeV from our UQM prediction\footnote{In the quenched limit, where the sea quark fluctuations are neglected, this difference becomes six times larger.}~\cite{Lu:2016mbb}.
Our prediction for the mass splitting of $\chi_{b1}(3P)$ and $\chi_{b0}(3P)$ is $23$ MeV, see Table~\ref{deltaM}. With the reference of the observed masses of the other two candidates of spin-triplet $3P$ bottomonium, this mass splitting helps us to predict precisely the mass of unknown $\chi_{b0}(3P)$ to be
\be
M[\chi_{b0}(3P)]=(10490\pm 3)~\mathrm{MeV}~.
\ee
The uncertainty in above prediction is calculated by taking the same percentage error [of $\mathcal O (10\%)$] in our mass splittings which we observed from CMS measurement. Our mass predictions respect the conventional pattern of splitting and support the standard mass hierarchy, where we have $M(\chi_{b2})>M(\chi_{b1})>M(\chi_{b0})$, which is in line with CMS measurement. A comparison of our UQM mass splittings with other quenched quark model predictions is given in Table~\ref{deltaM}.

\begin{table*}[!htbp]
\renewcommand\arraystretch{2}
  \centering
\begin{tabular}{cccccc}
  \hline\hline
  Mass Splitting & Our UQM~\cite{Lu:2016mbb} & GI~\cite{Godfrey:2015dia} & Modified GI~\cite{Wang:2018rjg} & CQM~\cite{Segovia:2016xqb} & Exp~\cite{Sirunyan:2018dff} \\
  \hline
  $\chi_{b1}(3P)- \chi_{b0}(3P)$     & $23$   & $16$ & $14$ & $13$ & $-$ \\
  $\chi_{b2}(3P)- \chi_{b1}(3P)$     & $12$   & $12$ & $12$ & $9$  & $(10.6\pm 0.64 \pm 0.17)$ \\

  \hline\hline
\end{tabular}
\caption{Mass splitting (in MeV) of $3P$-wave bottomonia in our UQM~\cite{Lu:2016mbb}, Godfrey-Isgur (GI) model~\cite{Godfrey:2015dia}, Modified GI model~\cite{Wang:2018rjg}, and constituent quark model (CQM)~\cite{Segovia:2016xqb}. The later three models are regarded as quenched quark models.}
\label{deltaM}
\end{table*}

\section{Ultrafine Splitting in UQM}

  It is more informative if we study the mass splitting in a multiplet
  instead of the total mass shift caused by the intermediate meson loop.
  For the states quite below the threshold,
  there is an interesting phenomenon~\cite{Burns:2011fu}:
  the magnitude of the mass splitting is suppressed by the probability
  of the bottomonium core, $P_{b\bar{b}}$, if we turn on the meson loop.

  There is also a pictorial explanation for this.
  Since under the potential model,
  the mass splitting $\delta M_0$ originates from the fine splitting Hamiltonian $H_I$.
  Up to the first order perturbation,
  we have $\delta M_0=\bra{\psi}H_I\ket{\psi}$, where $\psi$ is the two-body
  wave function in the quenched potential model.
  Since one of the coupled-channel effects is the wave function renormalization:
  $\bra{\psi}{\psi}\rangle=P_{b\bar{b}}<1$,
  one would simply expect that the $\delta M_0$ will be suppressed by this probability.

  Moreover, due to the closeness of the spectrum of a multiplet,
  we expect that the $P_{b\bar{b}}$ of the states in a same multiplet are nearly the same,
  i.e., $\delta M_0$ are all suppressed by a same quantity,
  leaving the relation
  \be\label{splitRelation}
  \delta M_P\equiv\frac{1}{9}\Big[M(\chi_{b0})+3 \cdot M(\chi_{b1})+5\cdot M(\chi_{b2})\Big]-M(h_b)=0
  \ee
  intact, even if the coupled-channel effects are turned on.
  Due to the remarkably small $\delta M_P$, we refer it as ``ultrafine splitting''.
  In our calculation, however, due to the finite size of the constituent quark,
  which is reflected by the smeared delta term, $\tilde{\delta}(r)$, instead of the true Dirac term\footnote{Such a \textit{smearing} of the Dirac delta term incorporating the contact spin-spin interaction with a finite range $1/\sigma$ is essential to regularize the delta function~\cite{Barnes:2005pb}.} in
  the spin dependent potential
  \bal\label{finestructure}
  V_{s}(r)& = \frac{1}{m^2_b}\bigg[ \left(\frac{2\alpha_s}{r^3}-\frac{\lambda}{2 r}\right)\vect{L}\cdot \vect{S}
  +\frac{32\pi \alpha_s}{9}~\tilde{\delta}(r)~\vect{S}_b\cdot \vect{S}_{\bar{b}} \nonumber\\
  &+\frac{4\alpha_s}{r^3}\left(\frac{\vect{S}_b\cdot \vect{S}_{\bar{b}}}{3}
  +\frac{(\vect{S}_b\cdot\vect{r}) (\vect{S}_{\bar{b}}\cdot \vect{r})}{r^2}\right) \bigg]~,\\
  \tilde{\delta}(r)&\equiv \bigg(\frac{\sigma}{\sqrt{\pi}} \bigg)^3 e^{-\sigma^2 r^2}~,\nonumber
  \eal
  where $\alpha_s$ and $\lambda$ are strengths of the color Coulomb and linear confinement potentials, respectively,
  and $\sigma$ is related to the width of Gaussian smeared function, the $\delta M_P$ relation of Eq.~\eqref{splitRelation}
  is already violated a little bit under the potential model which can be seen from Table~\ref{deltaMp} (second column),
  where we also include the corresponding experimental values.
  We can also extract the threshold effects by taking the mass shift $\Delta M$ instead of $M$ in $\delta M_P$ calculations.
  The $\delta M_P$ values obtained in this way are also given in Table~\ref{deltaMp} (third column).

\begin{table}[!htbp]
\renewcommand\arraystretch{2}
  \centering
\begin{tabular}{cccc}
  \hline\hline
  Multiplet & UQM prediction & CCE contribution & Experiment~\cite{Patrignani:2016xqp}\\
  \hline
  $1P$     & $1.17$ &  $0.06$ & $0.57(88)$ \\
  $2P$     & $1.38$ &  $0.19$ & $0.44(1.31)$ \\
  $3P$     & $-0.39$ & $2.08$ & $-$  \\

  \hline\hline
\end{tabular}
\caption{Ultrafine splitting ($\delta M_P$ in MeV) for the $P$-wave bottomonia.
The second to fourth columns are our unquenched quark model prediction, contribution from the coupled-channel effects and experimental results, respectively.
The contribution from coupled-channel effects can be obtained by replacing the mass of $\chi_{bJ}(nP)$ by their mass shift $\Delta M$.
Note that our results of $M_0$ violate Eq.~\eqref{splitRelation} a bit due to finite size of the constituent quark, as discussed in the text.}
\label{deltaMp}
\end{table}

  We can see from Table~\ref{tab:massProb} that although the mass shift for the $P$-wave multiplets is around 50~MeV,
  the modification of Eq.~\eqref{splitRelation} is not very large,
  except $\delta M_P(3P)$ which is far larger than $\delta M_P(2P)$ and $\delta M_P(1P)$.
  A worth mentioning feature here is the hierarchy of these ultrafine splittings originated
  from the CCE (third column of Table~\ref{deltaMp}), viz.,
  \be
  \delta M_P(3P) > \delta M_P(2P) > \delta M_P(1P)~,
  \ee
  which highlights that the coupled-channel effects bring meson masses closer together with respect to their
  bare values~\cite{Burns:2011fu}.

  \begin{table*}[!htbp]
  \renewcommand\arraystretch{1.5}
  \setlength{\tabcolsep}{5pt}
    \centering
    \begin{tabular}{c|c|ccc|ccc|c}
      \hline\hline
      Channels & $\delta M_0$ & $\widetilde{P}_{b\bar{b}}$ & $(\widetilde{P}_{b\bar{b}}\times\delta M_0)$ & $\delta M$ &
      $\widetilde{P}_{b\bar{b}}$ & $(\widetilde{P}_{b\bar{b}}\times\delta M_0)$ & $\delta M$ & $\delta M_{\text{Exp}}$ \\
      \cline{3-8}
       & & \multicolumn{3}{c|}{GEM}&\multicolumn{3}{c|}{SHO}& \\
      \hline
  $\Upsilon(1S)-\eta_b(1S)$ & 65.5 & 98.7 & 64.7 & 64.7 & 98.7 & 64.7 & 64.7 & 62.3 \\
  $\Upsilon(2S)-\eta_b(2S)$ & 30.7 & 95.5 & 29.3 & 29.4 & 95.9 & 29.4 & 29.5 & 24.3 \\
  $\Upsilon(3S)-\eta_b(3S)$ & 23.4 & 89.0 & 20.8 & 20.7 & 91.1 & 21.3 & 21.3 & -- \\
  \hline
  $\chi_{b0}(1P)-h_b(1P)$ & -35.6 & 97.2 & -34.6 & -34.5 & 97.1 & -34.6 & -34.4 & -39.9 \\
  $\chi_{b1}(1P)-h_b(1P)$ & -6.3 & 97.0 & -6.1 & -6.0 & 97.0 & -6.1 & -6.0 & -6.5 \\
  $\chi_{b2}(1P)-h_b(1P)$ & 13.2 & 96.9 & 12.8 & 12.6 & 96.8 & 12.8 & 12.7 & 12.9 \\
  \hline
  $\chi_{b0}(2P)-h_b(2P)$ & -31.2 & 93.0 & -29.0 & -28.9 & 93.4 & -29.2 & -29.1 & -27.3 \\
  $\chi_{b1}(2P)-h_b(2P)$ & -5.4 & 92.5 & -5.0 & -4.9 & 93.0 & -5.0 & -5.0 & -4.3 \\
  $\chi_{b2}(2P)-h_b(2P)$ & 12.2 & 92.1 & 11.2 & 11.2 & 92.7 & 11.3 & 11.2 & 8.8 \\
  \hline
  $\chi_{b0}(3P)-h_b(3P)$ & -29.2 & 56.9 & -16.6 & -27.5 & 54.3 & -15.8 & -28.3 & -- \\
  $\chi_{b1}(3P)-h_b(3P)$ & -5.0 & 69.4 & -3.5 & -4.5 & 72.5 & -3.6 & -4.6 & -- \\
  $\chi_{b2}(3P)-h_b(3P)$ & 11.9 & -- & -- & 7.5 & -- & -- & 7.7 & -- \\
      \hline\hline
    \end{tabular}
    \caption{
      The mass splitting (in MeV) in a same $(n,L)$ multiplet,
      where $\delta M_0$, $\delta M$ and $\delta M_{\text{Exp}}$
      represent the mass splitting in potential model, coupled-channel model and experiment, respectively.
      The $\widetilde P_{b\bar{b}}$ (in $\%$) is the weighted average of the probability,
      which for $P$- and $S$-wave is $\widetilde{P}_{b\bar{b}}=P_{b\bar{b}}(\chi_{bJ})$ and $\widetilde{P}_{b\bar{b}}= \frac{1}{4}P_{b\bar{b}}(\Upsilon) + \frac{3}{4}P_{b\bar{b}}(\eta_b)$, respectively. The details of the mass splitting are given in Appendix~\ref{Swave}, and the absolute probabilities $P_{b\bar{b}}$ are given in Table~\ref{tab:massProb}. GEM and SHO stand for the Gaussian expansion method~\cite{Hiyama:2003cu}
      and simple harmonic oscillator approximation, respectively, to fit the numerical wave functions.
    }\label{tab:ratioMass}
  \end{table*}

  Since, for the $P$-wave states, no matter whether the threshold effects are considered or not,
  $h_b$ is not affected by the fine interaction, i.e. the $\delta M = 0$.
  Hence, the $\chi_{bJ}$'s mass splitting are purely due to the $P_{b\bar{b}}$ of each $\chi_{bJ}$.
  Therefore, the weighted probability of the bottomonium core, $\widetilde{P}_{b\bar{b}}$,
  for $\chi_{bJ}(nP)$ multiplets is simply defined as $\widetilde{P}_{b\bar{b}}=P_{b\bar{b}}(\chi_{bJ})$.
  The weighted average probability for the $S$-wave bottomonia is discussed in Appendix~\ref{Swave}.
  From the Table~\ref{tab:ratioMass}, we can see that although the ($\widetilde{P}_{b\bar{b}}\times\delta M_0$)
  and $\delta M$ originate differently; one from the potential model and the other purely from the
  coupled-channel effects, but they are approximately equal to each other.
  The only large deviation comes from $\chi_{bJ}(3P)$.

  As explained above,
  this overall suppression is based on the assumption that the $\widetilde{P}_{b\bar{b}}$ is the same (or approximately the same) for a multiplet.
  Indeed, from Table~\ref{tab:massProb} we can see that this is quite reasonable assumption for the states which are far below the threshold.
  But for the $\chi_{b0}(3P)$,
  the $\widetilde{P}_{b\bar{b}}$ is quite different from that of $\chi_{b1}(3P)$,
  so this overall suppression does not make sense anymore.
  As a consequence, one should expect relatively large deviation from the $\delta M_P$ relation,
  as can be seen from $\delta M_P(3P)$ in Table~\ref{deltaMp}.

  The reason for this peculiar $\widetilde{P}_{b\bar{b}}$ is that
  even though the mass of $h_b(3P)$ and $\chi_{b1}(3P)$ is larger than the $\chi_{b0}(3P)$,
  they do not couple to the channel $B\bar{B}$, and the next open channel $B\bar{B}^*$ is somewhat farther from them.
  A net effect is that the $\widetilde{P}_{b\bar{b}}$ of $\chi_{b1}(3P)$ is larger than that of $\chi_{b0}(3P)$,
  breaking the $\widetilde{P}_{b\bar{b}}$ closeness assumption.
  This strong coupling of $\chi_{b0}(3P)$ to $B\bar{B}$ is also reflected by the large mass shift caused by $B\bar{B}$
  which can be seen from Table~\ref{tab:massProb}.
  The observed mismatch between ($\widetilde{P}_{b\bar{b}}\times \delta M_0$) and $\delta M$
  for $\chi_{bJ}(3P)$ multiplet is a smoking gun of the threshold effects which are beyond the quark potential model.

Recently, Lebed and Swanson also pointed out the remarkable importance of the $P$-wave heavy quarkonia~\cite{Lebed:2017yme}.
For $1P$ and $2P$ charmonia, the ultrafine splitting is found to be astonishingly small.
They argued that the ultrafine splitting can be used to delve the exoticness of the observed structure in the given multiplet~\cite{Lebed:2017xih}.
According to their analysis~\cite{Lebed:2017yme}, the quantity $\delta M_{n, L=1,2,3,...}$ is found to be very small for any radial excitation $n$, both for the $b\bar{b}$ and $c\bar{c}$ sectors. The obtained constraint on the $\delta M_{n,L}$ value is
\be
\delta M_{n, L=0,1,2,...} \ll \Lambda_{\textrm{QCD}}~.
\ee
This conclusion follows from several theoretical formalisms which do not consider coupled-channel effects or long-distance light-quark contributions in terms of intermediate meson-meson coupling to bare quarkonium states.
As discussed above, the operators corresponding to ultrafine splitting involve spin-spin interactions which are suppressed by $1/m_Q^{2}$, the standard expansion parameter for the heavy quarkonium, where $m_Q$ is the mass of heavy quark.
According to our point of view the above maxima is much large for the ultrafine splitting of $P$-wave bottomonia, see Table~\ref{deltaMp} for experimental corroboration. The more tight constraint could be
\be
\delta M_{n, L=1,2,3,...} \lesssim \frac{\Lambda_{\textrm{QCD}}^{3}}{m_{Q}^2}~.
\label{Ourcontraint}
\ee
Since, quantitatively the $P$-wave excitation for the bottomonium is equal to $\Lambda_{\textrm{QCD}}$, which describes the emergence of the dynamical QCD scale in above relation. The $\delta M_{n,L}$ for the bottomonia with $L=1$ is expected to be of $\mathcal O (1~\mathrm{MeV})$, which can be verified from our analysis of Table~\ref{deltaMp}.

The reason why $\delta M_{n, L=1,2,3,...}$ is exactly zero in the
quark model is a consequence of the pure delta function nature of the
$\vect{S}_b\cdot \vect{S}_{\bar{b}}$ term of Eq.~(\ref{finestructure}),
which is a perturbative one gluon exchange effect.
The non-perturbative effects can make an additional contribution to this term,
so that it is no longer a pure delta function. This give rise to introduce
the smearing of the delta function in the quark models~\cite{Barnes:2005pb,Lebed:2017yme}.
However, one could use different non-perturbative forms for the spin-spin operator
that contributes to the ultrafine splitting.
For instance, the ultrafine splitting computed at next-to-next-to-next-to leading order (N$^3$LO)
\cite{Kiyo:2014uca} in nonrelativistic QCD (NRQCD)~\cite{Caswell:1985ui,Brambilla:1999xf} is
\be
\delta M_{n, L=1}=\frac{m_b C^4_F \alpha^5_{s}}{432 \pi(n+1)^3} (4n_l-N_c)~,
\ee
where $C_F$ is the color factor of bottomonium, $n_l$ being the number of
light fermion species appearing in loop corrections, and $N_c$ is the number
of colors in QCD. The computed $\delta M_{n, L=1}$ values using NRQCD for the bottomonium
(with $m_b=4.5$ GeV and $\alpha_s (m_b)=0.2$) are;
$\delta M_{1P}=3.77$ keV, $\delta M_{2P}=1.12$ keV, and $\delta M_{3P}=0.47$ keV~\cite{Lebed:2017yme}.
The remarkable smallness of these values strengthen the constraint on the
$\delta M_{n, L=1,2,3,...}$ values presented in Eq.~(\ref{Ourcontraint}).
However, these NRQCD predictions are much smaller as compared to our UQM predictions
and corresponding experimental values, see Table~\ref{deltaMp}.
In conclusion, whatever the non-perturbative form for the spin-spin operator
is used, the $\delta M_{n, L=1}$ should be very small, hence satisfying the relation
of Eq.~(\ref{Ourcontraint}) quantitatively.

\section{Radiative Transitions}
\label{trans}

Radiative transitions of higher bottomonia are of considerable interest, since they can shed light on their internal structure and provide one of the few pathways between different $b\bar{b}$ multiplets. Particularly, for those states which can not directly produce at $e^+ e^-$ colliders (such as $P$-wave bottomonia), the radiative transitions serve as an elegant probe to explore such systems. In the quark model, the electric dipole ($E1$) transitions can be expressed as~\cite{Kwong:1988ae,Li:2009nr}
\be
\Gamma(n^{2S+1}L_J\to n'^{2S'+1}L'_{J'}+\gamma)=\frac{4}{3}C_{fi}\delta_{SS'}e^2_{b}\alpha |\bra{\psi_f}r\ket{\psi_i}|^2E^3_{\gamma},
\label{e1Decay}
\ee
where $e_b=-\frac{1}{3}$ is the $b$-quark charge, $\alpha$ is the fine structure constant, and $ E_\gamma$ denotes the energy of the emitted photon.
The spatial matrix elements $\bra{\psi_f}r\ket{\psi_i}$ involve the initial and final radial wave functions, and $C_{fi}$ are the angular matrix elements. They are represented as
\ba
\bra{\psi_f}r\ket{\psi_i}&=&\int_0^\infty R_f(r)R_i(r) r^3 dr, \label{rOverlap} \\
C_{fi}&=&\max(L,L')(2J'+1)
\left\{
\begin{array}{ccc}
L'& J'& S\\
J & L &1
\end{array}
\right\}^2.
\ea
The matrix elements $\bra{\psi_f}r\ket{\psi_i}$ are obtained numerically; for further details, we refer our studies~\cite{Lu:2016mbb,Lu:2017yhl}.
From Eq.~(\ref{rOverlap}),
we know that the value of the decay width depends on the details of the wave functions,
which are highly model dependent.
A model independent prediction can be achieved by focusing on the following decay ratios
\be
\Gamma \big(\chi_{bJ}(mP)\to \Upsilon(nS)+\gamma\big) \big/ \Gamma \big(\chi_{b0}(mP)\to \Upsilon(nS)+\gamma \big)~.
\ee
Since, in the quark model, the spatial wave function is the same for the states in the same multiplet.


From the above discussion, we know that the meson loop renormalizes the bottomnium wave function.
When the channel is above the corresponding open-bottom threshold (such as $B\bar{B}$ here),
the wave function cannot be normalized to $1$, this is still an open problem
(see e.g. Ref.~\cite{Kalashnikova:2005ui}).
On the other hand, the $B\bar{B}$ loop is still there, and have some CCE (such as mass renormalization).
We make the assumption that for the states above threshold (such as $\chi_{b2}(3P)$ here),
these open channels contribute equally to the wave functions of all $\chi_{bJ}(3P)$ states.
In fact this is a reasonable assumption, since we can see this from the Table~\ref{tab:massProb},
the probability of $B\bar{B}$ is vanishingly small ($0.31\%$ and $0.89\%$, less than $1\%$)
for both $\chi_{b0}(3P)$ and $\chi_{b1}(3P)$.


With the latest CMS data~\cite{Sirunyan:2018dff} and the $P_{b\bar{b}}$ in Table~\ref{tab:massProb},
our predictions of radiative decay ratios are listed in Table~\ref{tab:ratios}.
From the Table~\ref{tab:massProb}, one can see that the small $P_{b\bar{b}}[\chi_{b0}(3P)]$
make the ratios in the last three rows notably larger than that of the potential model predictions,
a peculiar feature of coupled-channel effects which can be tested in the upcoming experiments.

\begin{table}[!htbp]
\renewcommand\arraystretch{1.2}
 \centering
  \begin{tabular}{c|r@{ : }c@{ : }l|c@{ : }c@{ : }l}
   \hline\hline
   \diagbox{Decay Channel}{$\chi_{b0} : \chi_{b1} : \chi_{b2}$}{Model} &
   \multicolumn{3}{c|}{Potential Model} & \multicolumn{3}{c}{Unquenched Quark Model}\\
  \hline
  $\chi_{bJ}(1P) \to \Upsilon(1S)+\gamma$ & 1 & 3.80 & 7.20 & 1 & 3.79 & 7.18 \\
  \hline
  $\chi_{bJ}(2P) \to \Upsilon(1S)+\gamma$ & 1 & 3.27 & 5.71 & 1 & 3.25 & 5.65 \\
  $\chi_{bJ}(2P) \to \Upsilon(2S)+\gamma$ & 1 & 4.09 & 8.02 & 1 & 4.07 & 7.95 \\
  \hline
  $\chi_{bJ}(3P) \to \Upsilon(1S)+\gamma$ & 1 & 3.20 & 5.49 & 1 & 3.90 & 6.71 \\
  $\chi_{bJ}(3P) \to \Upsilon(2S)+\gamma$ & 1 & 3.46 & 6.15 & 1 & 4.22 & 7.51 \\
  $\chi_{bJ}(3P) \to \Upsilon(3S)+\gamma$ & 1 & 4.83 & 9.77 & 1 & 5.89 & 11.9 \\
  \hline\hline
  \end{tabular}
 \caption{
   Prediction for the ratios $\Gamma \big(\chi_{bJ}(mP)\to \Upsilon(nS)+\gamma \big) \big/ \Gamma \big(\chi_{b0}(mP)\to \Upsilon(nS)+\gamma \big)$. For potential model calculations, the parameters and quenched Hamiltonian are same as Ref.~\cite{Lu:2016mbb}.}
 \label{tab:ratios}
\end{table}

Another worth noting result from Table~\ref{tab:ratios} is the relative size of the ratios for
$\chi_{b0}(3P)$, which from the coupled-channel calculations is roughly $1:6:12$. This reflects that the
$\chi_{b0}(3P)$ has negligible radiative decay branching fraction with comparison to $\chi_{b1}(3P)$ and $\chi_{b2}(3P)$.
Compared with the potential model, the suppression of the $\chi_{b0}(3P)$'s radiative width in the UQM is
more consistent with the non-observation of the $\chi_{b0}(3P)$ in the recent
CMS search of $\chi_{bJ}(3P)\to \Upsilon(3S) \gamma$~\cite{Sirunyan:2018dff}.
This indicates that our UQM predictions are more reliable than the naive quark potential models.

\section{Conclusions}

The recent CMS study successfully distinguishs $\chi_{b1}(3P)$ and $\chi_{b2}(3P)$ for the first time, and measures their mass splitting which differs only $1$ MeV from our unquenched quark model predictions. This measurement gives us confidence to predict mass of the lowest candidate of $3P$ multiplet to be $M[\chi_{b0}(3P)]=(10490\pm 3)~\mathrm{MeV}$, based on our unquenched quark model results of the mass splittings of this multiplet.
We also analyze the ultrafine splittings of $P$-wave bottomonia up to $n=3$ in the framework of UQM, and put a constraint on them based on recent experimental corroboration. No matter which non-perturbative form for the spin-spin operator
is used, the ultrafine splitting for the $P$-wave bottomonia should be very small.
This analysis leads us to conclude that the coupled-channel effects play a crucial role to understand
the higher bottomonia close to open-flavor thresholds.

At last, we predict here to some extent model-independent ratios of the radiative decays of $\chi_{bJ}(nP)$ candidates. A worth mentioning observation is that the coupled-channel effects can enhance the radiative decay ratios of $\chi_{bJ}(3P)$ as compared to the naive potential model predictions.
The relative branching fraction of $\chi_{b0}(3P) \to \Upsilon(3S) \gamma$ is negligible as compared to the other candidates of this multiplet, which naturally explains its non-observation in recent CMS search.

We hope above highlighted features of coupled-channel model provide useful references for the understanding of higher $P$-wave bottomonia and can be explored in ongoing and future experiments.

\section*{Acknowledgements}

We are grateful to Timothy J. Burns, Feng-Kun Guo, Richard F. Lebed, and Thomas Mehen
for useful discussions and suggestions, and to Christoph Hanhart for careful read of this
manuscript and valuable remarks.
This work is supported
in part by the DFG (Grant No. TRR110) and the NSFC (Grant
No. 11621131001) through the funds provided to the Sino-German
CRC 110 ``Symmetries and the Emergence of Structure in QCD'',
and by the CAS-TWAS President's Fellowship
for International Ph.D. Students.

\begin{appendix}

\section{Bare Hamiltonian}
\label{bare}

Bare states are obtained by solving the Schr\"odinger
equation with the well-known Cornell potential~\cite{Eichten:1978tg,Eichten:1979ms}, which incorporates a spin-independent color Coulomb plus
linear confined (scalar) potential.
In the quenched limit, the potential can be written as
\be
V(r)=-\frac{4}{3} \frac{\alpha}{r}+\lambda r+c,
\ee
where $\alpha, \lambda$ and $c$ stand for the strength of color Coulomb potential,
the strength of linear confinement and mass renormalization, respectively.
The hyperfine and fine structures are generated by the spin dependent interactions
 \bal
  V_{s}(r)&= \frac{1}{m^2_b}\bigg[ \left(\frac{2\alpha_s}{r^3}-\frac{\lambda}{2 r}\right)\vect{L}\cdot \vect{S}
  +\frac{32\pi \alpha_s}{9}~\tilde{\delta}(r)~\vect{S}_b\cdot \vect{S}_{\bar{b}}\\ \nonumber
  &+\frac{4\alpha_s}{r^3}\left(\frac{\vect{S}_b\cdot \vect{S}_{\bar{b}}}{3}
  +\frac{(\vect{S}_b\cdot\vect{r}) (\vect{S}_{\bar{b}}\cdot \vect{r})}{r^2}\right) \bigg]~,
  \eal
where $\vect{L}$ denotes the relative orbital angular momentum,
$\vect{S}=\vect{S}_b+\vect{S}_{\bar{b}}$  is the total spin of the charm quark pairs
and $m_b$ is the bottom quark mass.
The smeared $\tilde{\delta}(r)$ function can be read from Eq.~(\ref{finestructure}) or Refs.~\cite{Barnes:2005pb,Li:2009ad}.
These spin dependent terms are treated as perturbations.

The Hamiltonian of the Schr\"{o}dinger equation in the quenched limit is represented as
\be\label{quenched}
H_0 =2m_b+\frac{p^2}{m_b}+V(r)+V_{s}(r).
\ee
The spatial wave functions and bare mass $M_0$ are obtained by
solving the Schr\"{o}dinger equation numerically using the Numerov method~\cite{Numerov:1927}.
The full bare-mass spectrum is given in Ref.~\cite{Lu:2016mbb}.

\section{Details of the Coupled-Channel Effects}
\label{prob}

As sketched by Fig.~\ref{ccDiagram}, the experimentally observed state
should be a mixture of pure quarkonium state (bare state) and $B$ meson continuum.
The coupled-channel effects can be deduced by following way
\ba
  H_0 \ket{\psi_0} &=& M_0 \ket {\psi_0} \\
  H_0 \ket{BC;p} &=& 0 \\
  H_{BC} \ket{\psi_0} &=& 0\\
  H_{BC} \ket{BC;p} &=& E_{BC}\ket{BC;p}\\
  H \ket{A}&=& M \ket{A},\label{Eigen}
\ea
where $M_0$ is the bare mass of the bottomonium and can be solved directly from Schr\"{o}dinger equation,
and $M$ is the physical mass.
The interaction between $B$ mesons is neglected.
When Eq.~(\ref{Eigen}) is projected onto each component, we immediately get
\be
\bra{\psi_0} H \ket{\psi}=c_0 M=c_0 M_0+ \int d^3p \, c_{BC}(p) \bra {\psi_0} H_I \ket{BC;p}, \label{eqnC2}
\ee
\be
\bra{BC;p} H \ket{\psi}=c_{BC}(p)M=c_{BC}(p) E_{BC}+c_0 \bra{BC;p} H_I \ket{\psi_0}. \label{eqnC4}
\ee
Solve $c_{BC}$ from Eq.~(\ref{eqnC4}), 
substitute back to Eq.~(\ref{eqnC2}) and eliminate the $c_0$ on both sides,
we get a integral equation
\be
M=M_0+\Delta M, \label{intEqn}
\ee
where $\Delta M$ is given in Eq.~(\ref{massShift}).
Once $M$ is solved, the coefficient of different components can be worked out either.
For states below threshold, the normalization condition $\ket{A}$ can be rewritten as
\be
|c_0|^2+\int d^3p|c_{BC}|^2=1
\ee
after the substitution of $c_{BC}$, we get the probability of the $b\bar{b}$ component.
The sum of $BC$ is restricted to the ground state $B_{(s)}$ mesons,
i.e. $B\bar{B}, B\bar{B}^{*}+h.c., B^{*}\bar{B}^{*}, B_s\bar{B}_s, B_s\bar{B}^{*}_s+h.c., B^{*}_s\bar{B}^{*}_s$.

The \cce/ calculation cannot proceed if the wave functions of the $\ket{\psi_0}$ and $BC$ components are not settled in Eq.(7).
Since the major part of the \cce/ calculation is encoded in the wave function overlap integration,
\bal\label{overlap}
 \bra{BC;p} H_I \ket{\psi_0} &= \int d^3k
\phi_0(\vec{k}+\vec{p}) \phi_B^*(\vec{k}+x \vec{p})\phi_C^*(\vec{k}+x \vec{p}) \nonumber\\
& ~~~~\times |\vec{k}| Y_1^m(\theta_{\vec{k}},\phi_{\vec{k}})~,
\eal
where $x=m_q/(m_Q+m_q)$, and $m_Q$ and $m_q$ denote the bottom quark and the light quark mass, respectively.
The $\phi_0, \phi_B$ and $\phi_C$ are the wave functions of $\ket{\psi_0}$ and $BC$ components,
respectively and the notation $*$ stands for the complex conjugate.
These wave functions are in momentum space, and they are obtained by the Fourier transformation
of the eigenfunctions of the bare Hamiltonian $H_0$.
More details can be found in our earlier works~\cite{Lu:2016mbb,Lu:2017yhl}.

\section{Ultrafine Mass Splitting for $S$-Wave Bottomonia}
\label{Swave}

  For the $S$-wave ($\eta_b$ and $\Upsilon$) bottomonia,
  we define
  \be
  \delta M_S\equiv \frac{32\pi \alpha}{9m_b^2} |R(0)|^2
  \ee
  Due to the $\vect{S}\cdot \vect{S}$ interaction term in Eq.~\eqref{finestructure},
  we have $\delta M_0$:
  \bal
  \delta M_0(\eta_b)&=-\frac{3}{4} \delta M_S~, \nonumber \\
  \delta M_0(\Upsilon)&=+\frac{1}{4} \delta M_S~.
  \eal
  After the suppression of $P_{b\bar{b}}({\eta_b})$ and $P_{b\bar{b}}({\Upsilon})$,
  the mass splitting becomes,
  \be
  M(\Upsilon)-M(\eta_b)\equiv \delta M(\Upsilon)-\delta M(\eta_b)= \bigg(\frac{1}{4}P_{b\bar{b}}(\Upsilon)+\frac{3}{4}P_{b\bar{b}}(\eta_b) \bigg)~\delta M_S~.
  \ee
  So for the $S$-wave bottomonium, we defined the weighted average of the $P_{b\bar{b}}$
  \be
  \widetilde{P}_{b\bar{b}}=\frac{1}{4}P_{b\bar{b}}(\Upsilon) + \frac{3}{4}P_{b\bar{b}}(\eta_b)~.
  \ee

\end{appendix}

\end{document}